\documentclass[10pt,twocolumn,superscriptaddress,english,10pt,prl,showpacs,floatfix,aps]{revtex4-2}

\usepackage[latin9]{inputenc}
\usepackage{amsthm}
\usepackage{amsmath}
\usepackage{amssymb}
\usepackage{esint}
\usepackage{graphicx}
\usepackage{upgreek}
\usepackage{siunitx}

\makeatletter

\@ifundefined{textcolor}{}
{%
 \definecolor{BLACK}{gray}{0}
 \definecolor{WHITE}{gray}{1}
 \definecolor{RED}{rgb}{1,0,0}
 \definecolor{GREEN}{rgb}{0,1,0}
 \definecolor{BLUE}{rgb}{0,0,1}
 \definecolor{CYAN}{cmyk}{1,0,0,0}
 \definecolor{MAGENTA}{cmyk}{0,1,0,0}
 \definecolor{YELLOW}{cmyk}{0,0,1,0}
}

\usepackage{soul}
\usepackage{xspace}
\usepackage{braket}
\usepackage[backref=true,
bookmarksnumbered=true,
bookmarks=true,
bookmarksopen=true,
colorlinks=true,
citecolor=blue,
linkcolor=blue,
anchorcolor=green,
urlcolor=blue,unicode=false]{hyperref}

\renewcommand{\v}[1]{\ensuremath{\mathbf{#1}}} 
 
\let\baraccent=\= 
\renewcommand{\=}[1]{\stackrel{#1}{=}} 

\newcommand{\mos}{$\text{MoS}_\text{2}~$}
\newcommand{\mosol}{$\text{MoS}_\text{2}$}

\newcommand{\didv}{d$I$/d$V$}
\newcommand{\molf}{Ethyl-Diaminodicyanoquinone\xspace} 
\newcommand{\mol}{ethyl-DADQ\xspace} 
\newcommand{\posg}{imidazolidine group\xspace} 
\newcommand{\negg}{dicyanomethylene group\xspace} 

\makeatother

\usepackage{babel}

\begin{document}

\title{Variations of vibronic states in densely-packed structures of molecules with intramolecular dipoles}

\author{Sergey Trishin}
\affiliation{\mbox{Fachbereich Physik, Freie Universit\"at Berlin, 14195 Berlin, Germany}}

\author{Christian Lotze}
\affiliation{\mbox{Fachbereich Physik, Freie Universit\"at Berlin, 14195 Berlin, Germany}}

\author{Johanna Richter}
\affiliation{\mbox{Fachbereich Physik, Freie Universit\"at Berlin, 14195 Berlin, Germany}}

\author{Ga\"el Reecht}
\affiliation{\mbox{Fachbereich Physik, Freie Universit\"at Berlin, 14195 Berlin, Germany}}

\author{Nils Krane}
\affiliation{\mbox{Fachbereich Physik, Freie Universit\"at Berlin, 14195 Berlin, Germany}}

\author{Philipp Rietsch}
\affiliation{\mbox{Institut f\"ur Chemie und Biochemie, Freie Universit\"at Berlin, 14195 Berlin, Germany}}

\author{Siegfried Eigler}
\affiliation{\mbox{Institut f\"ur Chemie und Biochemie, Freie Universit\"at Berlin, 14195 Berlin, Germany}}

\author{Katharina J. Franke}
\affiliation{\mbox{Fachbereich Physik, Freie Universit\"at Berlin, 14195 Berlin, Germany}}
\email{franke@physik.fu-berlin.de}


\begin{abstract}
Electrostatic potentials strongly affect molecular energy levels and charge states, providing the fascinating opportunity of molecular gating. Their influence on molecular vibrations remains less explored. Here, we investigate \molf molecules on a monolayer of \mos on Au(111) using scanning tunneling and atomic force microscopy and spectroscopy. These molecules exhibit a large dipole moment in gas phase, which we find to (partially) persist on the \mos monolayer. The self-assembled structures consist of chains, where the dipoles of neighboring molecules are aligned anti-parallel. Thanks to the decoupling efficiency of the molecular states from the metal by the \mos interlayer, we resolve vibronic states of the molecules, which vary in intensity depending on the molecular surrounding. We suggest that the vibrations are strongly damped by electrostatic interactions with the environment.
\end{abstract}


\maketitle

\section{Introduction}

Molecules adsorbed on surfaces constitute a versatile platform for the design of hybrid organic-inorganic devices \cite{Xiang2016, Goronzy2018}. The energy-level alignment is largely determined by the interactions between the molecule and the substrate due to hybridization, charge transfer and screening \cite{Braun2009, Otero2017, Torrente2008a}. These effects are particularly strong when the molecules are in direct contact to a metal substrate, often leading to the loss of molecular function and tunability. To maintain the molecular properties, inorganic insulating or semiconducting interlayers have been inserted between the molecules and metal substrate \cite{Qiu2004, Repp2005a, Garnica2013,Schulz2013, Krane2018, Yang2018}. The decoupling is also beneficial for spectroscopic investigations of the molecular states, as the excitations are long-lived. The corresponding narrow linewidths then allow for the resolution of densely-spaced spectroscopic features, such as vibronic states \cite{Qiu2004, Pradhan2005, Ogawa2007, Matino2011, Franke2012, Schulz2013, Mehler2018, Krane2018}. 

Although the interlayers decouple the molecules from the substrate, there may still be charge transfer depending on the alignment of the energy levels of the molecules with respect to the Fermi level of the underlying metal \cite{Xiang2016, Willenbockel2015}. Control of the charge state is highly desirable for devices and may be achieved by a gate electrode \cite{Riss2014, Wickenburg2016}. However, the implementation of a gate electrode is not always feasible in experiment. Other strategies rely on the control of the charge state by self-assembly of donor and acceptor molecules \cite{Torrente2008b}, ideally on the decoupling layers \cite{Kumar2021}. Recently, it has been shown that the presence of a molecular dipole can be used for tuning the energy level alignment of neighboring molecules and, thus, gain control over the charge state \cite{Homberg2020, Li2022}. This approach also offers the unique opportunity to investigate the effect of local electrostatic potentials on the molecular properties. One interesting aspect is the modification of molecular vibrations in the presence of an electric field. It has been found that approaching an STM tip to a molecule leads to a shift of the vibrational modes \cite{Vitali2010, Okabayashi2018, Homberg2022a}. However, in this case, it is difficult to disentangle the influence of chemical forces from the electrostatic forces. 

Here, we aim at resolving the effect of electrostatic forces on molecular vibrations by control of the molecular environment. For this, we chose a molecule with a large intrinsic dipole moment to create a structure with non-negligible electrostatic potential. \molf (\mol) consists of two charge-separated moieties - an electronegative dicyanomethylene group and an electropositive imidazolidine group (Figure\,\ref{fig1}d) - which result in a large dipole moment of around 17.3 Debye in gas phase \cite{Rietsch2019}. 
Surprisingly, this dipole moment was found to be enlarged when the molecule is in contact to a metal substrate due to charge transfer \cite{Trishin2022}. The interaction with the substrate prohibited the resolution of vibronic states. Hence, we now use a monolayer of \mos for decoupling the molecules from the metal substrate. We find that the molecules self-assemble in quasi-one dimensional arrangements and preserve a dipole moment. The decoupling efficiency allows us to  detect vibronic states, which we find to be strongly suppressed in densely packed molecular chains. We speculate that intermolecular electrostatic interactions lead to a significant damping of the vibrations.

\section{Results and discussion}

\subsection{Self-assembly of \mol molecules on \mos}
\begin{figure*}[ht]
\includegraphics[width=0.9\textwidth]{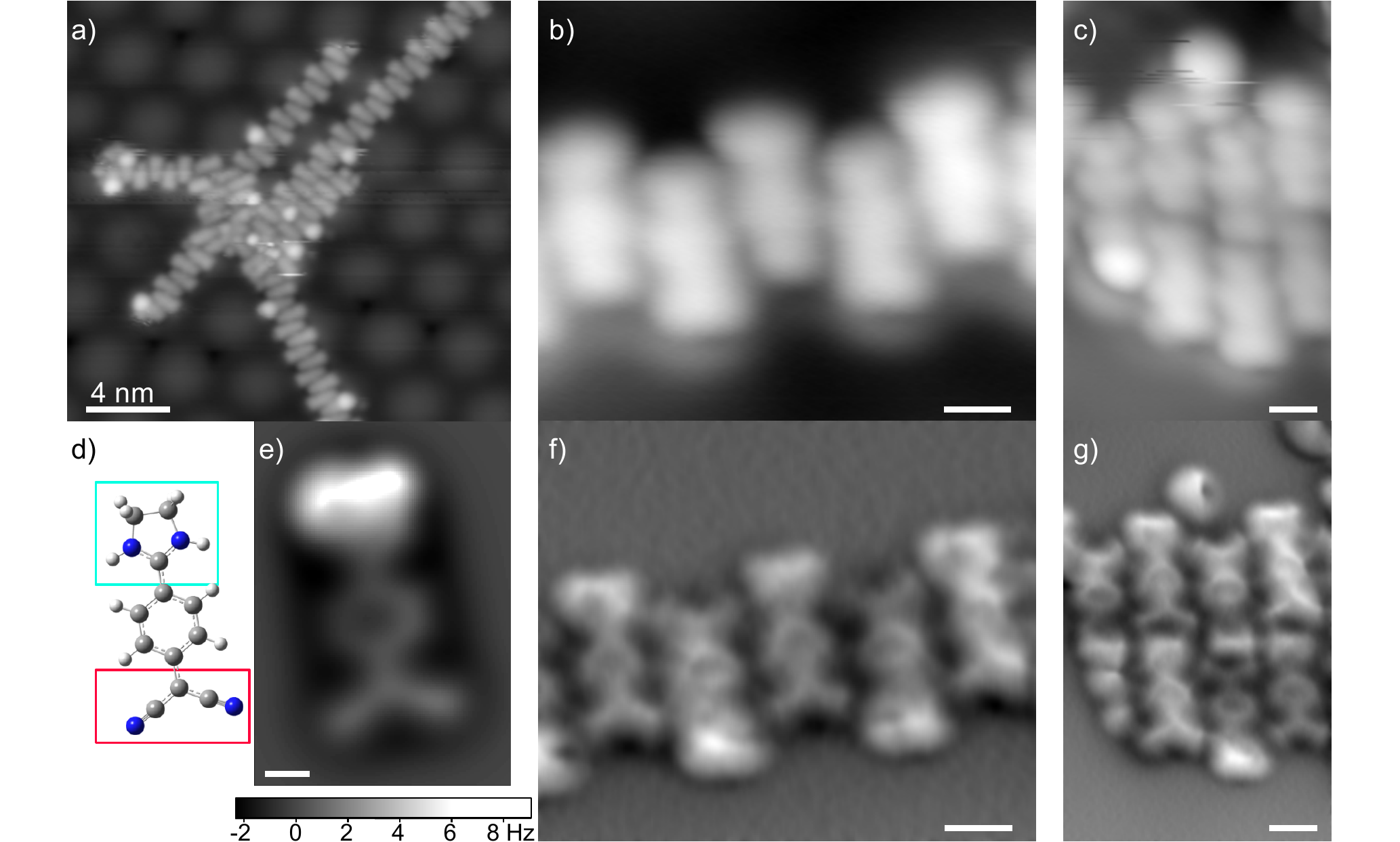}
\caption{\label{fig1}
  a) STM topography of \mol adsorbed on a monolayer of \mosol/Au(111). b) STM image taken with a Cl functionalized tip of a single-stranded molecular chain. c) STM image taken with a Cl functionalized tip of a double-stranded molecular chain. d) Stick-and-ball model of the \mol molecule. It consists of a benzene ring, which has two moieties bound to it, a \posg (top, teal) and a \negg (bottom, red).  
  e) The molecule has been placed on a graphene sheet for mimicking the van der Waals interactions with the surface. Based on this structure, we simulated the frequency-shift signal of the \mol when probed at constant-height with tip-sample distance of 6.9 \,\AA\ employing the probe particle model \cite{Hapala2014,Krejc2017}. The legend below the map gives the $\Delta f$ signal in units of Hz. The map is to scale with (d). 
  f-g) $\Delta f$ images recorded on the same area as the STM images in (b) and (c). The STM topographies in (a) and (c) were recorded at a setpoint of 1\,V, 30\,pA, the one in (b) at  520\,mV, 30\,pA. The constant-height images in (f) and (g) were recorded at 1\,V, 30\,pA. After switching the feedback off, the bias voltage was set to 0\,V and the tip approached by 0.1\,\AA\ to the substrate. The scale bars in (b),(c),(f),(g) are 5\,\AA\ and the one in (e) is 2\, \AA. 
  }
  \end{figure*}
Deposition of \mol molecules at 230\,K on a monolayer of \mos on a Au(111) substrate leads to the self-assembly in quasi-one-dimensional arrangements (Figure\,\ref{fig1}a). We refer to these molecular structures as chains in the following. The chains follow a preferred orientation with respect to the underlying moir\'e structure, which forms as a result of a lattice mismatch between the terminating Au layer and the \mos and is seen as a hexagonal superstructure with a periodicity of 3.3\,nm in the background \cite{Krane2018a, Bana2018}.
At low molecular coverage, we find that most chains only consist of a single molecular row, while double stranded chains or even wider chains remain scarce. We discuss the single- and double-stranded chains in the following.  

 We first analyze the structure of the self-assembled single-stranded chains. These chains consist of a zig-zag arrangement of the molecules, with the individual molecules imaged as an almost oval shape with one termination being slightly thinner than the other one. A similar arrangement has been reported for the same molecules in direct contact to a Au(111) surface \cite{Trishin2022}. A Cl functionalized tip (see Methods) enhances the asymmetric appearance of the individual molecules (Figure \,\ref{fig1}b) and evidences the alternating anti-parallel orientation with respect to the nearest neighbours. While the STM image is not conclusive on the orientation of the individual molecules, frequency-shift ($\Delta f$) images of the same area allow for an unambiguous identification of the molecular arrangement (Figure\,\ref{fig1}f). The observed shape directly reflects the atomic structure of the molecule (cp. Figure\,\ref{fig1}d) with the dicyanomethylene-, benzene ring and imidazolidine group lying flat on the surface. The largest elevation is found at the position of the H atoms at the imidazolidine termination. The experimental images are well reproduced by simulations based on the probe-particle model by Hapala et al. \cite{Hapala2014}, when the molecules are placed parallel to the surface and the H atoms at the imidazoline unit point towards the tip (Figure\,\ref{fig1}e). This detailed insight into the structure reveals that the cyano groups face the imidazolidine group, supporting CN--HN hydrogen bonds, which contribute to the stabilization of the structure.
 
  Using the same Cl functionalized tip, we also investigated the double-stranded chains. The STM and AFM images (Figure \,\ref{fig1}c,g) reveal a similar zig-zag pattern of alternating molecular arrangements along the chains, while the molecules arrange head-to-tail across the chains.

\subsection{Intramolecular dipole}
\begin{figure}[!t]
\includegraphics[width=0.45\textwidth]{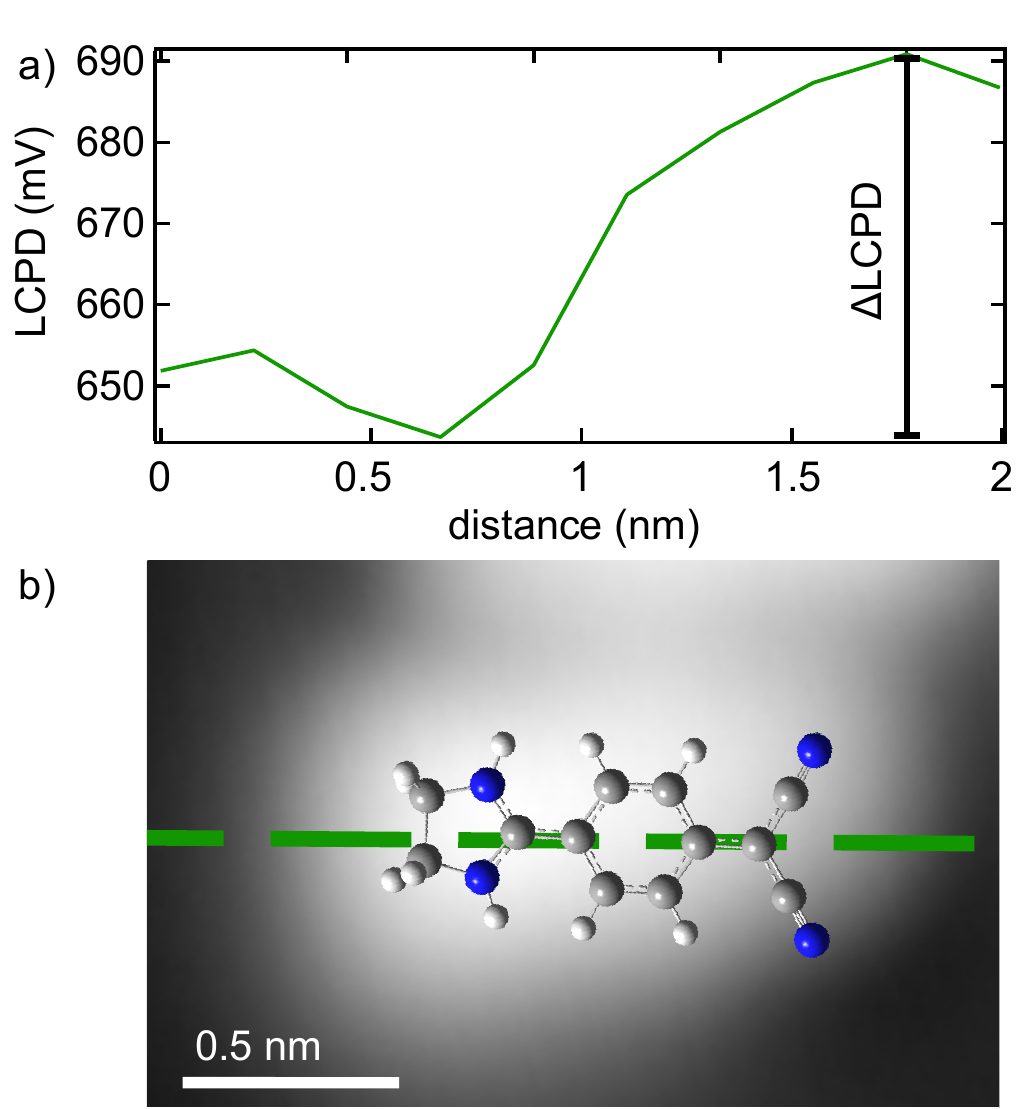}
\caption{\label{F:lcpd}
  a) Extracted LCPD signal over a line of densely spaced spectra, as indicated with the green dashed line.  The spectra were taken at a setpoint of -1.6\,V, 400\,pA.
  b) STM topography of a molecule at the edge of a one-molecule wide chain. A stick-and-ball structure model was overlaid for clarity. The topography was recorded at a setpoint of 1\,V, 30\,pA.}
\end{figure}

The alternating structure and head-to-tail arrangement of the molecules suggests that electrostatic interactions may also contribute to stabilizing the arrangements assuming that the molecular dipole moment is conserved on the surface.  
To find out if an intramolecular dipole persists, we mapped the local-contact-potential-difference (LCPD) across the molecules, which is obtained from the maxima of the voltage-dependent frequency-shift ($\Delta f-V$) curves \cite{Gross2009a,Mohn2012}. 
Figure \ref{F:lcpd}a shows the extracted LCPD values from a line of spaced spectra along a single \mol located at the end of a single-stranded chain (see Figure \ref{F:lcpd}b). The lowest LCPD value is found above the imidazolidine group.  In contrast, the LCPD value increases over the molecule towards the dicyanomethylene group, where it peaks. This distribution is up to our expectations, as the imidazolidine is the positively charged moiety of the molecule, whereas the dicyanomethylene group is negatively charged. The experiments thus let us conclude on the presence of an intramolecular dipole of the \mol on the \mos layer. Unfortunately, we cannot assess the magnitude of the dipole moment from the experiment. We note that the same molecules in direct contact to Au(111) also revealed the presence of an intramolecular dipole, which was shown to be enhanced as compared to the gas-phase molecule by charge transfer from the substrate \cite{Trishin2022}.

\subsection{Molecular energy levels and vibronic states in single-stranded chains} 

\begin{figure}[!t]
\includegraphics[width=0.45\textwidth]{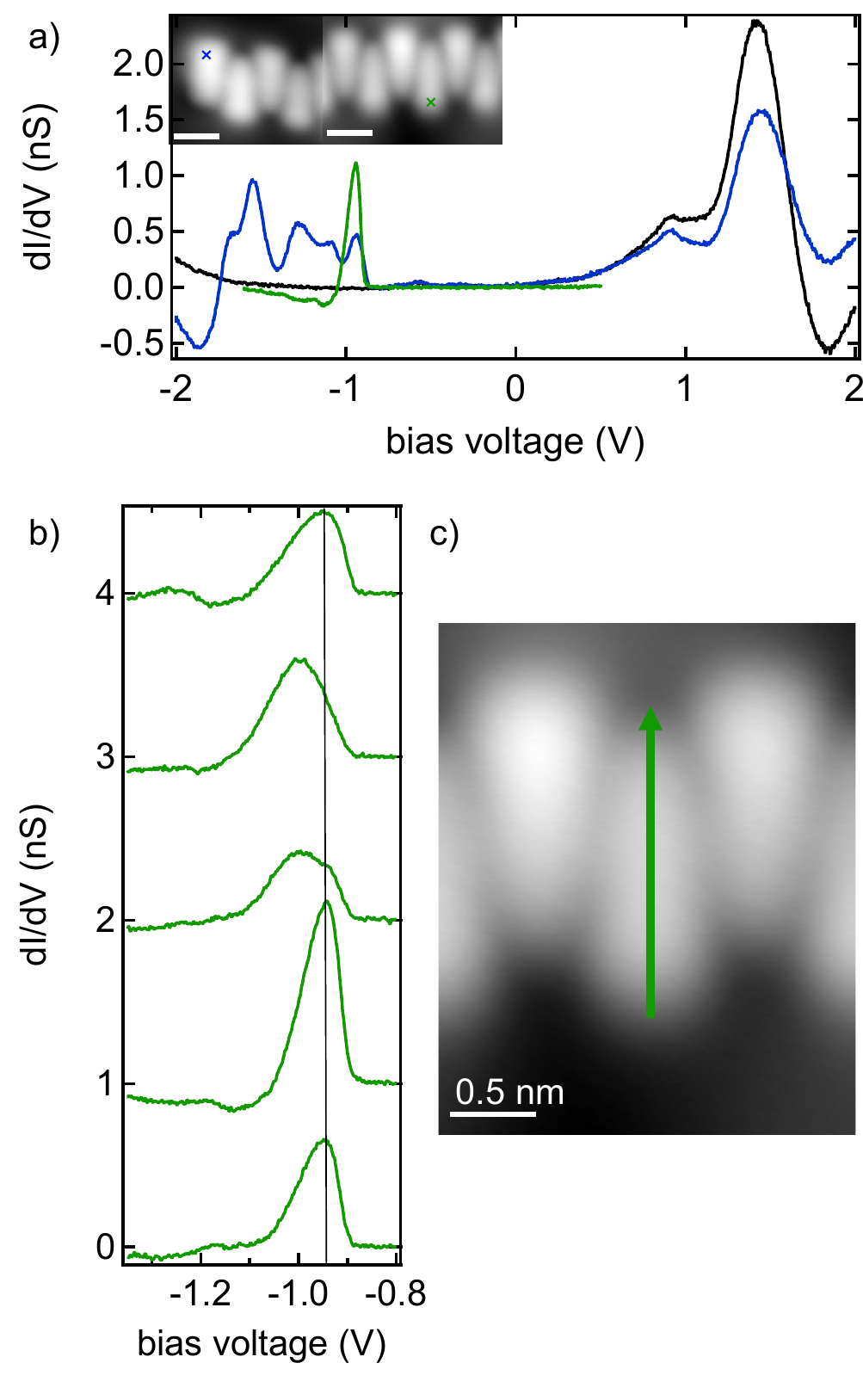}
\caption{\label{F:estates}
  a) \didv spectra recorded on the monolayer-island of \mos on Au(111) shown in black. \didv spectra performed on the \posg of a single \mol at the end of a molecular chain (blue) and a molecule inside a chain (green) with the locations shown in the inset. Several resonances appear inside the electronic bandgap of the \mosol, which we attribute to tunnelling into the HOMO of the molecule and vibronic levels associated with it. The spectra were recorded at setpoints of 2\,V, 1\,nA (black), -1.6\,V, 400\,pA (blue) and -1.6\,V, 50\,pA (green). The scale bars are 1\,nm.
  b) \didv spectra performed along the green line indicated in (c). The resonances vary along the molecule as described in the text. The spectra were recorded at a setpoint of -1.6\,V, 50\,pA and offset by 1\,nS for clarity. c) STM topography of a molecular chain. All topographies were recorded at a setpoint of 1\,V, 30\,pA.
  }
\end{figure}

Next, we turn to the investigation of the electronic structure of the \mol on \mosol/Au(111). The monolayer of \mos itself exhibits a band gap around the Fermi level, where the onset of the conduction band is found at $\sim$ 0.5\,V and the valence band maximum is located at $\sim$-1.4\,V (black spectrum in Figure \ref{F:estates}a) \cite{Miwa2014, Krane2018a}. The valence and conduction band edges are smeared out due to hybridization of the S layer with the Au substrate state \cite{Bruix2016, Krane2016}.

Spectra recorded on the \mol molecules (Figure\,\ref{F:estates}a blue and green) show additional peaks at negative bias voltages inside the semiconducting gap, corresponding to excitations of a positive ion resonance. The resonance structure (onset at $\sim$-0.9\,V) is much sharper than observed for the same molecules in direct contact to the Au(111) substrate \cite{Trishin2022}. The smaller linewidth indicates longer excited state lifetimes and less hybridization with the substrate. These observations are in agreement with earlier studies of organic molecules on \mosol/Au(111), which pointed out the decoupling efficiency of the monolayer \mos \cite{Krane2018,Reecht2020,Yousofnejad2020}.

The onset of the resonance is followed by a set of additional peaks, which are very pronounced on the molecule at a chains' end (blue spectrum in Figure\,\ref{F:estates}a) and less obvious on molecules embedded inside the bulk of a chain (green spectrum in Figure\,\ref{F:estates}a and Figure\,\ref{F:estates}b). Sidebands to the ion resonances are typically associated to vibronic states \cite{Qiu2004, Pradhan2005, Ogawa2007, Matino2011, Franke2012, Schulz2013, Mehler2018}. The strongest peaks originate from the excitation of vibrations with large Huang-Rhys factors, which represent the electron-phonon coupling strength. 
The strong suppression of the vibronic satellite peaks on molecules surrounded by neighbors suggest a significant damping of the vibrational states \cite{Repp2005, Fatayer2018}. We speculate that the dipolar environment strongly affects molecular vibrations, in particular those with dipolar character. 

\begin{figure}
    \centering
    \includegraphics[width=0.45\textwidth]{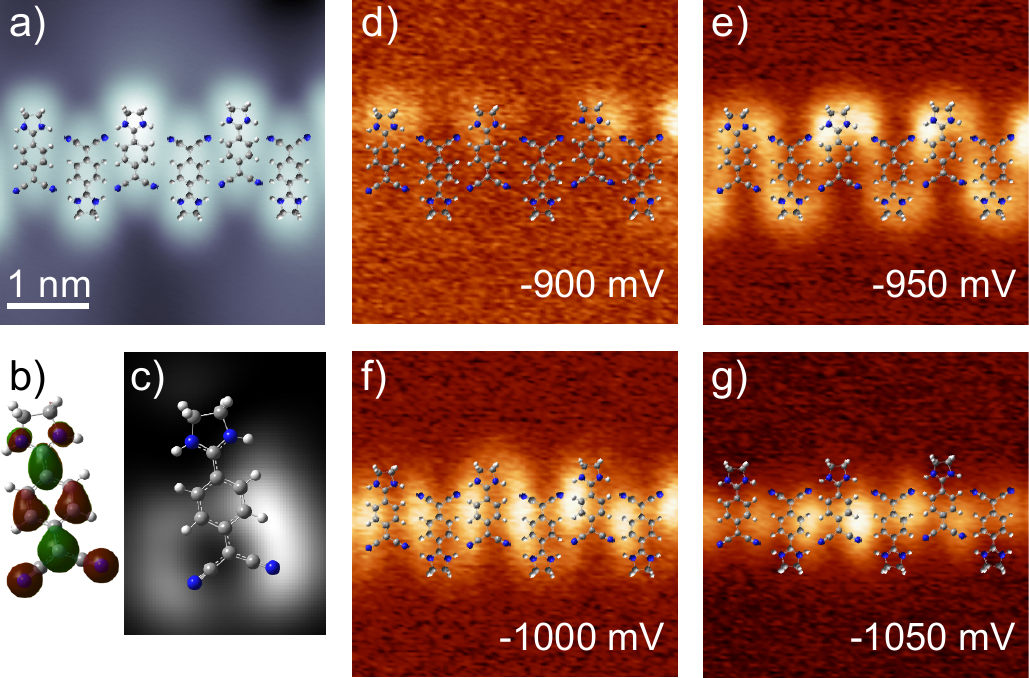}
    \caption{a) STM topography of a single-stranded chain of \mol molecules. b) DFT simulation of the HOMO of the \mol placed on a sheet of graphene (see Methods for details). c) Simulated Tersoff-Hamann constant-height image of the HOMO using an s-wave tip at a distance of 5 \AA\ above the molecule. The most pronounced shape is U-like around the dicyanomethylene group. d)-g) Constant-height \didv maps taken at the same location as the topography in (a). The energies of the maps are indicated in the bottom-right corner. All images were taken at an initial setpoint of 1\,V, 30\,pA before the feedback was switched off, the tip was additionally approached towards the sample by 2\,\AA. }
    \label{F:maps}
\end{figure}

We also observe a local variation of the vibronic spectra along the individual molecules (Figure\,\ref{F:estates}b,c). The spectra on the imidazolidine group show a sharp onset of a peak at -950\,mV, with a broader high-energy shoulder. The asymmetric lineshape suggests the presence of at least one vibronic satellite peak. At the center of the molecule the intensity ratio between shoulder and peak is reversed, corroborating the interpretation of a second peak in the first spectra. The intensity ratio changes again at the dicyanomethylene termination. 
To map out these changes in intensity ratio, we show constant-height \didv maps at the peak energies in Figure\,\ref{F:maps}(d-g).  While the off-resonance map at -900\,mV shows the largest intensity on the imidazolidine units, the first map on the molecular resonance at -950\,mV shows additional intensity next to the dicyanomethylene group, which gains strength at -1000\,mV. The map at -1050\,mV remains with the largest intensity between the molecules. To understand these intensity variations, we compare the maps to the theoretical electronic structure of the \mol molecule.

Excitation of a positive ion resonance in tunneling spectroscopy implies probing the occupied molecular orbitals. We calculated the highest occupied molecular orbital (HOMO) of the \mol molecule using density functional theory (see Methods, Figure\,\ref{F:maps}b). When the flat molecule is probed with an s-wave tip at a certain height, the Tersoff-Hamann model \cite{Tersoff1985} predicts the HOMO to appear with a U shape centered on the benzene ring and opening around the dicyanomethylene group (Figure\,\ref{F:maps}c).

For comparison to the experimental maps, we first need to eliminate the influence of the topographic height of the \mol on the surface. The off-resonance map at -900\,mV (Figure\,\ref{F:maps}d) revealed intensity at the imidazolidine group, indicating an elevated height in agreement with the AFM images in Figure\,\ref{fig1}. 
Neglecting the intensity at the imidazolidine group, the remaining intensity around the dicyanomethylene group on the resonance (at -950\,mV) resembles the Tersoff-Hamann map. We thus ascribe the first resonance within the \mos gap to tunneling through the HOMO. The very different intensity distribution at higher energies does not agree with the shape of the HOMO. As the HOMO-1 is found separated by 1.3\,eV from the HOMO in gas-phase calculations, the different shape cannot be explained by excitation of a lower-lying orbital. Instead, we suggest that vibration-assisted tunneling is responsible for the increased intensity at locations, where the HOMO itself exhibits a nodal plane. Inelastic excitation of an out-of-plane mode changes the wavefunction overlap with the tip. In case of an asymmetric mode around a nodal plane, an increase in the tunneling matrix element leads to a new vibration-assisted tunneling channel \cite{Pavlicek2013, Reecht2020}. This model -together with the topographic influence- could account for the variation of peak heights in Figure\,\ref{F:estates}b and the deviation from plain orbital shapes in Figure\,\ref{F:maps}(e-g).

Moreover, we recall that the spectra varied significantly between molecules embedded in the chain and at the chains's termination (Figure\,\ref{F:estates}a). This indicates that the molecular environment affects the molecular vibrations and, thus, also the vibration-assisted tunneling.

\subsection{Vibronic states in double-stranded molecular chains}

\begin{figure}
    \centering
    \includegraphics[width=0.49\textwidth]{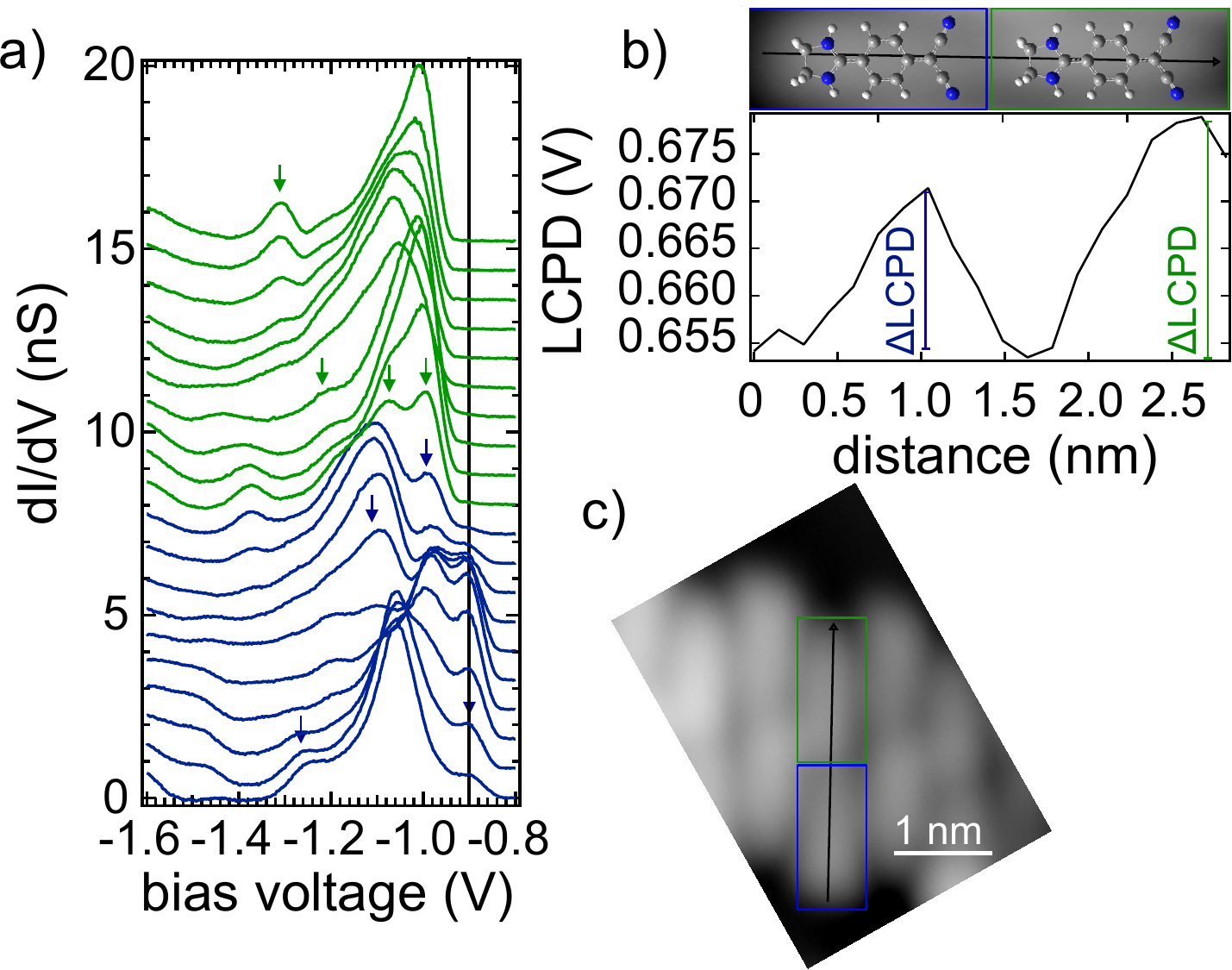}
    \caption{a) \didv spectra around the first positive ion resonance performed along molecules of a double-stranded chain, as indicated in (c). The spectra are offset for clarity. The arrows indicate the position of the most pronounced resonances. Blue and green spectra correspond to the blue and green boxed molecule in (c). The spectra are offset by 0.8\,nS for clarity.  b) LCPD values extracted from measurements at the same positions as those in (a) and shown above the graph. c) STM topography of a double-stranded chain. The coloured boxes indicate the two probed molecules in (a) and (b), which have a different electrostatic environment. STM images were recorded with a setpoint of 1\,V, 30\,pA, the spectra were recorded at a setpoint of -1.6\,V, 100\,pA with the tip being additionally approached towards the sample by 0.6\,\AA.}
    \label{F:spectra2}
\end{figure}

To unravel the influence of the electronic environment we investigated the vibronic sidebands to the HOMO-derived resonance of the \mol molecules in the double-stranded chains (Figure \ref{F:spectra2}). As shown above, the molecular orientation along these chains is similar to the one in the single-stranded chains with a second row added in parallel. 
This assembly imposes different electronic environments to the molecules within the chain. Half of the molecules have the \posg sticking out of the chain (indicated with the blue box in \ref{F:spectra2}c) and the other half of the molecules are embedded more inside the chain (indicated with the green box).

To resolve how this structure affects the electronic properties of the molecules, we recorded a line of densely spaced spectra along a molecular row, as indicated with the black arrow in Figure\,\ref{F:spectra2}c. The results are shown in Figure\,\ref{F:spectra2}a with the color code identifying the corresponding boxed molecules in the topography. 

Describing first the blue spectra, one can see that several peaks emerge, which are marked with the blue arrows. Mainly four different resonances can be resolved, at energies of around -900\,mV, -980\,mV, -1100\,mV and -1270\,mV, with their intensities varying along the molecule. 
Whereas the resonance at -900\,mV has a very low intensity at the imidazolidine group of the molecule and is strongest in the center, the resonances at  -1100\,mV and -1270\,mV have their main contribution to the signal at the terminations of the molecule. The resonance at -980\,mV is mainly located at the center and the \negg of the molecule.
The spectra taken along the green boxed molecule are not a simple repetition of the spectra. First, we note that the onset of the resonance structure is shifted by $\approx$90\,mV to higher energies. We then resolve again four resonances with varying intensity along the molecules (at -990\,mV, -1075\,mV, -1220\,mV and -1300\,mV). The overall shift can be explained by different screening from the environment due to different location of the molecule on the moir\'e structure and the different molecular neighborhood \cite{Torrente2008a}. However, the resonance sidebands are not simply subject to a rigid shift. The intensity variations again point to a significant damping of the molecular vibrations due to the different electrostatic environment, which is different for the two \mol molecules.

\begin{figure}
    \centering
    \includegraphics[width=0.49\textwidth]{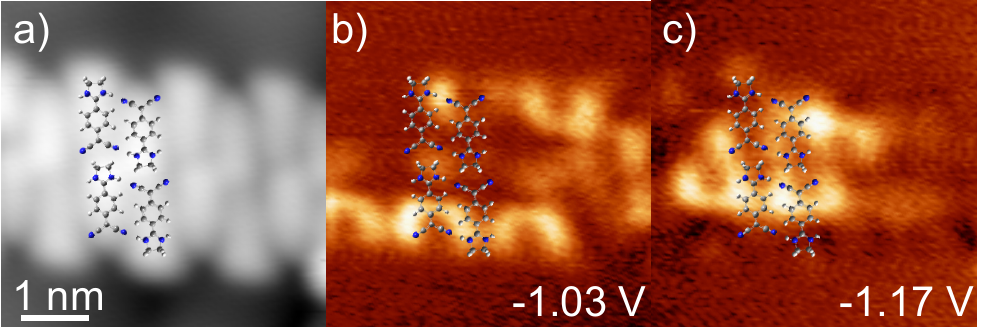}
    \caption{a) STM topography of a double-stranded chain of \mol molecules. A molecular model is superimposed on the image for better comparison. b),c) Constant-height \didv maps taken at the same location as the topography in (a). The energies of the maps are indicated in the bottom-right corner. All images were taken at an initial setpoint of 1\,V, 30\,pA, the tip was additionally approached towards the sample by 2\,\AA.}
    \label{F:maps2}
\end{figure}

 To probe the electrostatic environment, we present the LCPD along the same two molecules in Figure\,\ref{F:spectra2}b. The spatial variation along both molecules is in agreement with presence of an intramolecular dipole albeit with a different strength of the two molecules. The molecular environment is thus also expressed in different screening of the molecular dipole. The variations in vibronic damping may thus be correlated with the electrostatic potential imposed by the environment.

 In the last step, we again map out the molecular resonance structure for comparison with the vibronic spectra (Figure\,\ref{F:maps2}). The map at -1030\,mV shows some \mol molecules with the shape of the HOMO as expected from the Tersoff-Hamann simulations in Figure\,\ref{F:maps}c. In contrast to the single-stranded chains, these maps are thus not strongly influenced by an elevated imidazolidine group and, therefore, suggest a flatter adsorption configuration. The different onsets of the molecular resonances observed in Figure\,\ref{F:spectra2}a are reflected by some molecules still being dark at this bias voltage. At this particular bias voltage, the molecules with high intensity are those equivalent to the \mol molecules indicated in green. The maps thus clearly reflect the alternating electronic structure of the molecules sticking out of the chains compared to the ones embedded deeper in the chain.  At larger bias voltage, the \didv\ signal gains strength again between the molecules, similar to the earlier observation on the single-stranded chains. This again shows that vibration-assisted tunneling dominates the vibronic spectra.

\section{Conclusions}
We investigated organic molecules (\mol) on a monolayer of \mos on Au(111) using scanning tunneling and atomic force microscopy and spectroscopy. These molecules were chosen for their large intramolecular dipole moment which raises the question of electrostatic intermolecular interactions and their effect on the properties of the layer. Importantly, we first showed that an intramolecular dipole moment is preserved within the self-assembled molecular arrangements. The \mos decoupling layer then allowed us to resolve vibronic states of the individual molecules. Our main observation is a strong variation of the vibronic structure depending on the molecular surrounding. Our data indicates a strong damping of molecular vibrations when the molecule is affected by the electrostatic environment of neighboring \mol molecules. We also observe signatures of vibration-assisted tunneling being influenced by the surrounding. Our results suggest that the electrostatic environment can be used to tune the strength of damping and stiffening of molecular vibrations. Indeed, damping may be beneficial for the performance of  molecular-scale devices.

\section{Methods}
The Au(111) substrate was cleaned by repeated sputter-annealing cycles in an ultra-high vacuum chamber. The \mos islands were subsequently grown by depositing Mo atoms in H$_{2}$S gas of p = $2 \times 10^{-5}$ mbar and simultaneous annealing of the gold substrate to 800 K. 
The \mol molecules were deposited at a substrate temperature of 230\,K. The as-prepared sample was transferred into the scanning tunneling microscope. The microscope is equipped with a qPlus sensor allowing for simultaneous measurements of the tunneling current and frequency shift \cite{Giessibl2000}. The base temperature was 4.5\,K.

The LCPD value was determined from bias-dependent frequency-shift-spectra ($\Delta f$-V). The frequency shift depends quadratically on the applied bias voltage. The maximum of the inverse parabola corresponds to the LCPD. Its value is extracted from a fit. 

To image molecules at very short tip-molecule distances an in-situ functionalization of the tip is required. We attached chlorine atoms to the tip by picking them up from chlorinated Fe-octaethyl-porphyrin molecules co-deposited on the same substrate. Such a tip has been used for taking the STM and simultaneous AFM images in Figure\,\ref{fig1}b,c,f,g. Otherwise we used a Au coated tip for STM measurements and  \didv spectroscopy. 

DFT calculations of the \mol molecules were performed employing the Gaussian 16 A.03 package, using the 6-31++G(d,p) basis set and the B3PW91 functional \cite{Gaussian}. For structure optimization the molecule was placed on a graphene sheet applying the ONIOM method and using the UFF force field for the graphene atoms, to mimic the weak interaction with the surface \cite{DAPPRICH1999}. As a result, the molecular structure is flattened as compared to a gas-phase optimized single molecule (dihedral angle of $\approx 4^{\circ}$\ vs\ $\approx7^{\circ}$\ between the benzene and imidazolidine group.
This flattened structure shows similar orbital shapes and simulated STM Tersoff-Hamann images to the non-flat molecule but shows significantly better agreement of the simulated AFM $\Delta f$-images with the experimental ones.

\textbf{Acknowledgments}
We acknowledge financial support by the Deutsche Forschungsgemeinschaft (DFG)
through SFB~951 ``Hybrid Inorganic/Organic Systems for Opto-Electronics''
(project number 182087777, project A14).


%

\end{document}